\documentclass[a4paper,fleqn,usenatbib]{mnras}

\usepackage{newtxtext}
\usepackage{newtxmath}
\usepackage{natbib}

\usepackage[T1]{fontenc}
\usepackage{aecompl}

\usepackage{graphicx}				
\usepackage{bm}						
\usepackage{amsmath}				
\usepackage{amssymb}				
\usepackage{physics}				
\usepackage[nameinlink, capitalize]{cleveref}	
\usepackage{array}

\newcommand{\erad}{\theta_\mathrm{E}}		
\newcommand{\critd}{\Sigma_\mathrm{crit}}	
\newcommand{\reff}{r_\mathrm{eff}}			
\newcommand{\imra}{{\theta_\mathrm{r}}}		
\newcommand{\flra}{{f_\mathrm{r}}}			
\newcommand{\valpha}{\bm{\alpha}}			
\newcommand{\vbeta}{\bm{\beta}}				
\newcommand{\vtheta}{\bm{\theta}}			
\newcommand{\posa}{\phi_\mathrm{L}}			
\newcommand{\poss}{\phi_\mathrm{S}}			
\newcommand{\pvf}{\bm{x}}					
\newcommand{\pvl}{\pvf_\mathrm{L}}			
\newcommand{\pvs}{\pvf_\mathrm{S}}			
\newcommand{\etheta}{\theta_\varepsilon}	
\newcommand{\sigmad}{\sigma_\mathrm{d}}		
\newcommand{\sg}{\sigma_\gamma}				
\newcommand{\prob}[2]{\mathrm{Pr}\!\left(#1\vert#2\right)}		
\newcommand{\prior}[1]{\mathrm{Pr}\!\left(#1\right)}			

\title[Mass profiles from lensing I]{Galaxy mass profiles from strong
  lensing I: The circular power-law model}

\author[C. M. O'Riordan et al.]{
C. M. O'Riordan,$^{1}$\thanks{E-mail: conor.oriordan15@imperial.ac.uk}
S. J. Warren$^{1}$ and
D. J. Mortlock$^{1,2,3}$
\\
$^{1}$Astrophysics Group, Blackett Laboratory, Imperial College London, London, SW7 2AZ, United Kingdom\\
$^{2}$Department of Mathematics, Imperial College London, London, SW7 2AZ, UK\\
$^{3}$Department of Astronomy, Stockholm University, Albanova, SE-10691 Stockholm, Sweden
}

\date{Accepted XXX. Received YYY; in original form ZZZ}

\pubyear{2019}

\begin{document}
\label{firstpage}
\pagerange{\pageref{firstpage}--\pageref{lastpage}}
\maketitle

\begin{abstract}

\noindent
In this series of papers we develop a formalism for constraining mass profiles in strong gravitational lenses with extended images, using fluxes in addition to positional information. We start in this paper with a circular power-law profile and show that the slope $\gamma$ is uniquely determined by only two observables: the flux ratio $f_1/f_2$ and the image position ratio $\theta_1/\theta_2$ of the two images. We derive an analytic expression relating these two observables to the slope, a result which does not depend on the Einstein angle or the structure or brightness of the source. We then find an expression for the uncertainty on the slope $\sigma_\gamma$ that depends only on the position ratio $\theta_1/\theta_2$ and the total S/N in the images. For example, in a system with position ratio $\theta_1/\theta_2=0.5$, S/N $=100$ and $\gamma=2$ we find that $\gamma$ is constrained to a precision of $\pm0.03$. We then test these results against a series of mock observations. We invert the images and fit an 11 parameter model, including ellipticity and position angle for both lens and source and measure the uncertainty on $\gamma$. We find agreement with the theoretical estimate for all mock observations. In future papers we will examine the radial range of the galaxy over which the constraint on the slope applies, and extend the analysis to elliptical lenses.
\end{abstract}

\begin{keywords}
gravitational lensing: strong
\end{keywords}



\section{Introduction}
\label{sec:intro}

Strong gravitational lensing, that is, when a source is multiply imaged by a lensing galaxy, can provide measurements of the masses of galaxies that are much more accurate than measurements from dynamics. Strong lensing gives, principally, the total projected mass interior to the Einstein angle $\erad$ (the radius inside which the average density is high enough for multiple imaging to occur). Unfortunately strong lensing, on its own, yields only limited information on the radial profile of the mass \---\ or at least that is the prevailing view. \citet{Kochanek1991} made the first detailed study of the use of strong lensing to measure mass profiles, drawing pessimistic conclusions. Dynamical measurements complement the information from lensing, and, as summarised below, this combination has provided a wealth of results on the mass profiles of galaxies, and their evolution. Two thorough reviews \citep{Kochanek2006, Treu2010a} cover the use of strong lensing for measuring galaxy mass profiles.

The accurate measurement of the mass profile of the lensing galaxy is central to the determination of the Hubble constant from measured time delays \citep{Birrer2018,Oguri2003,Rusu2017,Sluse2017,Sonnenfeld2018,Suyu2010,Suyu2017,Wong2017,Wong2018,Xu2016}. Another application is in the detection of mass substructure in galaxy haloes \citep{Hezaveh2016,Koopmans2005,Nierenberg2014}, where it is important to determine simultaneously the best-fit smooth galaxy mass profile, against which substructure is identified, to avoid biasing the substructure results. Strong lensing mass determinations have also proved useful in studies of the stellar initial mass function \citep{Auger2010,Brewer2014,Spiniello2011,Treu2010b}. The measurement of the exponent $\gamma$ of power-law fits to galaxy density profiles, $\rho(r) \propto r^{-\gamma}$, its mean value and scatter, and their evolution, has provided an exacting test for theories of the formation of galaxies \citep{Barnabe2009,Barnabe2011,Bolton2008a,Dye2005,Dye2008,Koopmans2003,Koopmans2006,Koopmans2009,Rusin2002,Spiniello2011,Treu2002,Treu2010b,Courteau2014}.

Most current methods to measure galaxy mass profiles of distant galaxies combine the lensing observables with dynamical information, ostensibly to break degeneracies which exist in each method alone. Pure dynamics methods suffer from a degeneracy between the measured mass profile and the radial profile of the anisotropy of the stellar orbits \citep[e.g.][]{Courteau2014,Koopmans2003}. Pure lensing methods also suffer from degeneracies in determining the mass profile. The best known of these is the mass-sheet transformation \citep[MST,][]{Falco1985}, but \citet[e.g.][]{Schneider2014} have identified a wider set of degeneracies, known collectively as the source-position transformation, of which the mass-sheet transformation is a particular example. In the original and most widely used lensing+dynamics analysis method \citep{Treu2002,Koopmans2003,Koopmans2006}, the only lensing information used is the total mass inside the Einstein angle, which breaks the dynamical mass-anisotropy degeneracy. This is sometimes framed the other way round, that the dynamics breaks the mass-sheet degeneracy. \citet{Barnabe2007} developed a more advanced lensing+dynamics method which combines 2D velocity dispersion information with the full lensing flux information. These lensing+dynamics studies have shown that total mass profiles in early-type galaxies are close to isothermal, i.e. $\gamma\sim 2$, and such a profile slope in a spherical mass distribution gives a flat rotation curve.

Returning to pure strong-lensing analyses, lensing degeneracies may often be avoided by the use of a parametric form for the mass profile, e.g. a power law \citep{Wagner2018}. The results of such a parametric analysis will be useful to the extent that the model is correct. The power-law profile is a popular model for parameterising the mass profile in the central regions of galaxies, and this has been the preferred model in lensing+dynamics analyses, as well as pure dynamics studies \citep[e.g.][]{Chae2014} and in simulations \citep[e.g.][]{Wang2018,Mukherjee2018,Schaller2015,Xu2016}.

Although there have been a number of pure lensing analyses that use power-law fits, and that provide good constraints on $\gamma$, rather little attention has been paid to the results. One reason we have identified that may explain this is a general distrust within the lensing community of the use of flux information to provide constraints on the mass profile. This is for historical reasons, because the majority of strong lensing theory was developed for analysing quasar images. Fits to multiply-imaged quasars only ever use the positions of the images, possibly supplemented by time delays, because the flux information is unreliable due to variability, microlensing or substructure. As a consequence the theory of the use of fluxes remains comparatively undeveloped. Without the fluxes there is too little information in images of lensed quasars to constrain the parameters of the mass profile, as demonstrated in the original analysis by \citet{Kochanek1991}. Improved constraints require positional information from multiple source systems, with $>4$ images \citep{Cohn2001,Trotter2000}, but such systems are rare.

For extended sources, where the source is a galaxy rather than a
quasar, variability and microlensing may be neglected, so the flux
information is useful, provided extinction in the lens may be
neglected i.e. the lens may be treated as transparent. Substructure remains an issue \---\ which is why it can be measured (see references above). The images can be highly stretched, providing multiple resolution elements, each yielding flux and position information. Analyses of mass profiles using extended sources \citep[e.g.][]{Rusin2002,Dye2005,Dye2008,Hezaveh2016,Bellagamba2017,Spingola2018,Wucknitz2004} have provided the most precise measurements of $\gamma,$ with uncertainties as small as $0.02$. This is substantially more precise than the original lensing+dynamics method and at least as precise as the updated method mentioned above. A direct comparison of the lensing+dynamics analysis of the Einstein ring 0047-2803 by \citet{Koopmans2003}, with a pure lensing analysis of the same data, was made by \citet{Dye2005}. They found that the pure lensing analysis provided much stronger constraints, and in particular (their fig. 3) it was possible to make an accurate decomposition using a two component stars+dark matter model, and to measure the stellar mass-to-light ratio, removing the strong degeneracies between the two components found in the corresponding lensing+dynamics analysis.

Given that strong gravitational lensing on its own can measure $\gamma$ as precisely as $0.02$, why then is it not the preferred method for measuring galaxy mass profiles? The reason seems to be the belief that the constraints on the slope only apply over the radial range of the images, and this is often only a small fraction of $\erad$. This belief is made explicit in \citet{Kochanek2006} which states `it is important to remember that the actual constraints on the density structure really only apply over the range of radii spanned by the lensed images', while \citet{Treu2010a} states `It should be noted that lensing is mostly sensitive to the projected mass-density slope at the location of the images, rather than the average inside the images'. Numerous other papers contain similar statements, or make clear that the measured value of $\gamma$ quoted only applies at the Einstein angle \citep[e.g.][]{Chae2014,Hezaveh2016,Koopmans2006,Spingola2018,Suyu2017,vandeVen2009}. The earliest example of this view that we can find is \citet{Kochanek1995} who states `As with all lens systems the mass distribution is strongly constrained only over the multiply imaged region'. We have not found similar earlier quotes, so this may be the origin of the idea. Nevertheless we have not found a calculation which explicitly proves this statement. We will examine this belief closely in later papers in this series.

To summarise, flux information has been underused in the analysis of strongly lensed images of extended sources, and the theory of the constraints on the galaxy mass profile provided by such images is underdeveloped. Our goal in these papers is to produce a detailed understanding of how images of extended sources constrain the mass profile, and to develop a method to determine the lens surface mass density, together with the uncertainties, as a function of radius \---\ in other words to determine where in the profile the constraints come from.

In this first paper we analyse the simple case of a transparent circularly symmetric power-law lens, and develop a general theory of how accurately the power-law slope $\gamma$ may be measured for any image configuration. We treat the case of an isolated galaxy. In general, and especially for high precision work, for example measuring the Hubble constant, contributions from other galaxies nearby or along the line of sight must also be accounted for. Fortunately good solutions to this problem exist: for example \citet{McCully2017} present a comprehensive treatment of this issue. A discussion of the circular power-law lens is provided by \citet{Kochanek2006}, but it is limited to consideration of positional information. The circular power-law lens is also addressed by \citet{Suyu2012}, although not in a systematic way. Treating the case of an extended source, and inverting a synthetic lensed image, she concludes that `the relative thickness of the arcs accurately constrains the lens mass distribution'. In their inversion of the lens SDSS J1148+1930, \citet{Bellagamba2017} draw a similar conclusion, identifying the radial magnification ratio between pairs of images as the relevant quantity \citep[see also][]{Sonnenfeld2018}. Here we consider the general two-image case for the circular power-law lens. We show how the positional information on its own, or the flux (i.e. magnification) information on its own, measures only a degenerate combination of $\gamma$ and the source position $\beta$, but that the combination of positional and flux information breaks the degeneracies and provides a measurement of both quantities. In later papers we consider more complicated mass models, and extend the analysis to include elliptical lenses.

In \cref{sec:theory} we present a theoretical analysis of the problem, and derive an expression that 
relates $\gamma$ to two observables; the ratio of the radial angle of the two images (the position ratio), and the ratio of the fluxes.
We go on to derive the uncertainty on the measured value of $\gamma$ as a function of the total signal to noise ratio of the two images and the position ratio. These results are independent of the structure of the source. In \cref{sec:method} and \cref{sec:results} 
we create realistic artificial observations for isolated, transparent, isothermal, circular lenses over a range of source positions, and constrain the parameters of a power-law model by inverting the images. The method
for creating these observations is detailed in \cref{sec:method}, and the results are compared to the theoretical estimates in \cref{sec:results}. \Cref{sec:correlations} lists expressions for the correlations between the parameters in the power-law model.

\section{The circular power-law lens}
\label{sec:theory}

The mass density of a singular spherical power-law lens is
\begin{equation}
	\rho(r) \propto r^{-\gamma},
\end{equation}
where $r$ is the physical three-dimensional radial coordinate and $\gamma$ is the mass profile slope. We adopt the thin lens approximation \citep{Schneider1992} which allows us to integrate the above into a two-dimensional surface-mass density
\begin{equation}
	\Sigma(R) \propto R^{1-\gamma},
\end{equation}
where $R$ is now a physical two-dimensional radius perpendicular to the line of sight. This is the singular circular power-law lens.We can then scale this distribution by the distances involved and work strictly in angular coordinates as follows. We define the dimensionless surface mass density $\kappa$, also called the convergence, as
\begin{equation}
\kappa(\theta) = \frac{\Sigma(D_\mathrm{d}\theta)}{\critd},
\end{equation}
where $\theta$ is the angular radial coordinate. The critical density of the system, $\critd$ is
\begin{equation}
	\critd=\frac{c^2}{4\pi G}\frac{D_\mathrm{s}}{D_\mathrm{ds} D_\mathrm{d}},
\end{equation}
where $D_\mathrm{ds}$, $D_\mathrm{d}$ and $D_\mathrm{s}$ are the angular diameter distances from the lens (also called the deflector) to the source, from the observer to the lens and from the observer to the source respectively \citep[e.g.][]{Schneider1992}.

\begin{figure}
	\includegraphics[width=0.5\textwidth]{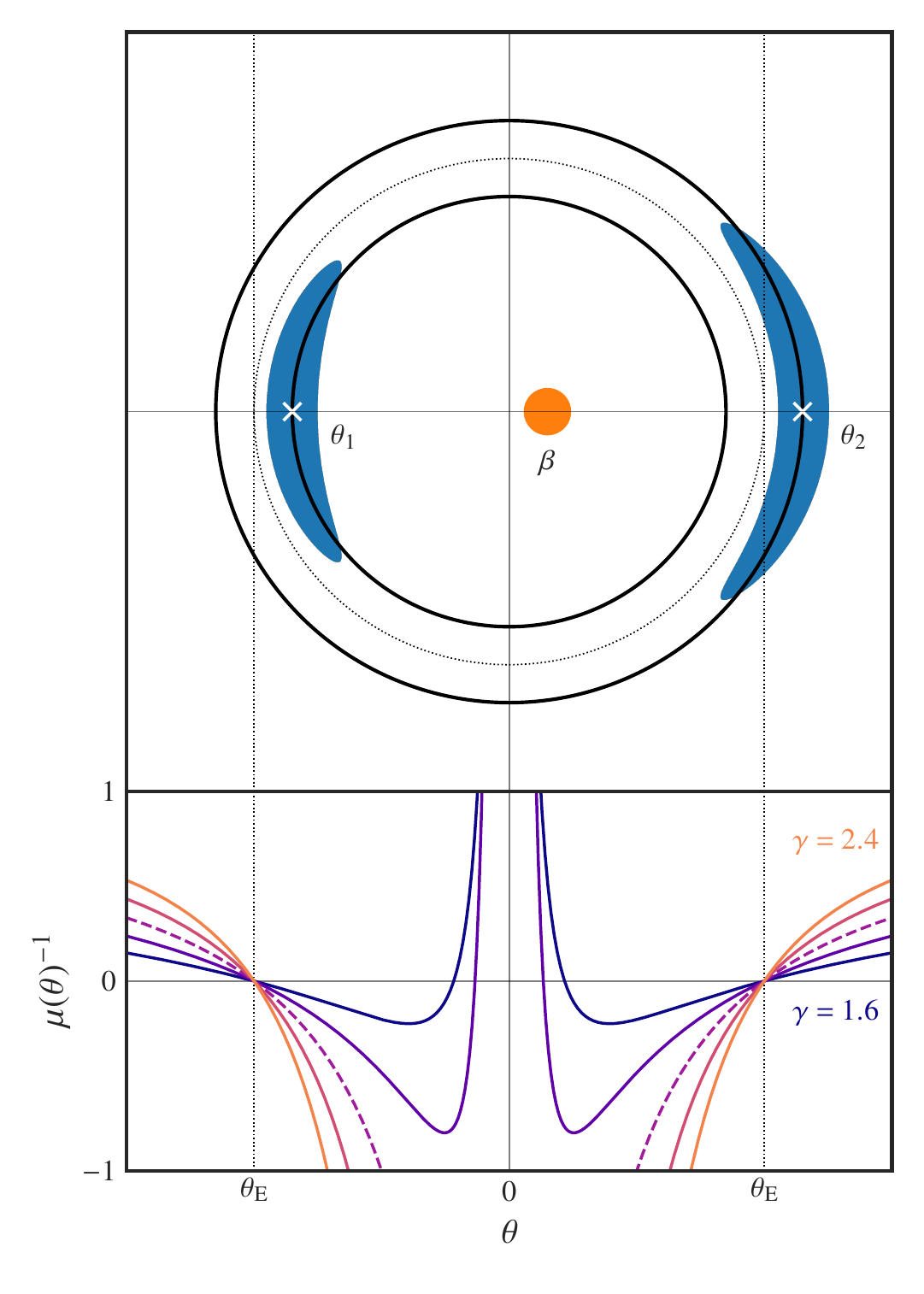}
	\caption{\label{fig:image-structure} The image structure in our example system (upper frame) with the inverse magnification for different mass profile slopes (lower frame). The dashed line shows $\mu(\theta)^{-1}$ for $\gamma=2$ with the other lines spaced by $\Delta\gamma=0.2$. The magnification diverges at $\theta=\erad$ for all profiles and again towards the centre for profiles with $\gamma<2$. The blue shaded areas are the images for a disc source (shown in orange at $\beta$) with $\gamma=2$.}
\end{figure}
The total mass $M$ enclosed by a radius $\theta$ is
\begin{equation}
M(\theta) = \critd D_\mathrm{d}^2 \int_{0}^{\theta} \kappa(\theta')2\pi \theta'\dd\theta'. 
\end{equation}
The Einstein angle, $\erad$, is the angular radius such that
\begin{equation}
	M(\erad)=\critd\pi\erad^2 D_\mathrm{d}^2,
\end{equation}
or in other words, the radius inside which the average density is $\critd$. The appropriately normalised form of $\kappa$ is then
\begin{equation}
\kappa(\theta) = \frac{3 - \gamma}{2}\left(\frac{\erad}{\theta}\right)^{\gamma-1}.
\end{equation}
This power-law profile includes three special cases: $\gamma=1$, $M(\theta)\propto\theta^2$ and the distribution is a sheet of uniform density; $\gamma=2$, $M(\theta)\propto\theta$, corresponding to an isothermal sphere; $\gamma=3$, $M(\theta)\propto1$, equivalent to the lens being a point mass.

The lens system is illustrated in \cref{fig:image-structure} and is characterised by four parameters: the mass profile slope $\gamma$, the Einstein angle $\erad$, the source position $\beta$ and the source flux $f_\mathrm{S}$. We make the simplifying assumption that the source size is small relative to $\erad$. Then we can derive image positions and magnifications without reference to the structure of the source. We also assume that the centroid of the mass distribution is known.

For $\gamma>2$ there are always two images. For $\gamma<2$ the number of images is either one or three. Third central (demagnified) images are a rarity in lensing observations \citep[e.g.][]{Rusin2001} and we do not take them into account here. In any case, the presence of a third image would only enhance the constraints on $\gamma$ as it provides more information in the image plane while also restricting the model to profiles with $\gamma<2$. Our analysis is therefore limited to the use of two images. The image plane then provides four observable quantities: the two image positions $\theta_1$ (inner) and $\theta_2$ (outer) and the flux of each image $f_1$ and $f_2$. With four observables, the four system parameters may be constrained directly. 

In the following we show that a measurement of the ratios $\flra=f_1/f_2$ and $\imra=\theta_1/\theta_2$ is sufficient to determine the mass profile slope $\gamma$, independent of the other system parameters. We then proceed to derive the uncertainty on this measurement as a function of $\imra$ and the signal to noise ratio of the observation.

\subsection{Image positions}

In the following the angles $\beta$ (source position), $\theta$ (image position), and $\alpha$ (reduced deflection angle) are always positive quantities. For a source at some position $\beta$, the lens forms images at $\theta_1$ and $\theta_2$ where $\theta_1<\erad<\theta_2$. The positions themselves are given by the roots of the lens equation
\begin{equation}
	\label{eq:lens-eqn}
	\left|\theta-\alpha(\theta)\right|-\beta=0\:,
\end{equation}
and $\alpha(\theta)$ for the power-law profile is given by
\begin{equation}
	\label{eq:deflection-angle}
	\alpha(\theta) = \erad\left(\frac{\erad}{\theta}\right)^{\gamma-2}.
\end{equation}
Using the lens equation, and the fact that both images share the same
source position, as shown by \citet{Kochanek2006} we can write the Einstein angle as a function only of $\gamma$ and the image positions, yielding
\begin{equation}
	\label{eq:einstein-radius}
	\erad^{\gamma-1} = \frac{\theta_1 + \theta_2}{\theta_1^{2-\gamma} + \theta_2^{2-\gamma}}.
\end{equation}
We could also have eliminated $\erad$ rather than $\beta$ when combining \cref{eq:lens-eqn,eq:deflection-angle}. In this case one obtains
\begin{equation}
	\label{eq:source-position}
	\beta = \frac{\theta_2^{\gamma-1} - \theta_1^{\gamma-1}}{\theta_1^{\gamma-2} + \theta_2^{\gamma-2}}.
\end{equation}
We define the reduced source position, $\beta'=\beta/\theta_2$. Combining this with \cref{eq:source-position} we get an expression for $\beta'$ in terms of $\gamma$ and $\imra$:
\begin{equation}
	\label{eq:beta-dashed}
	\beta'=\frac{1-\imra^{\gamma-1}}{1+\imra^{\gamma-2}}.
\end{equation}
We choose $\beta'=\beta/\theta_2$ as the definition of the reduced source position rather than the natural $\beta/\erad$, because the resulting expression \cref{eq:beta-dashed} is much simpler.

\subsection{Image fluxes}
\label{sec:image-fluxes}

The flux at a given image position is the product of the source flux and the magnification at that position. The flux ratio is then
\begin{equation}
	\label{eq:flra}
	\flra = \left|\frac{\mu(\theta_1)}{\mu(\theta_2)}\right|,
\end{equation}
where $\mu(\theta)$ is the scalar magnification, provided below. The flux ratio is independent of the source flux. We take the absolute value of the ratio, because the magnification, defined by
\begin{equation}
	\label{eq:magnification}
\mu(\theta)^{-1}=\left[1-\left(\frac{\erad}{\theta}\right)^{\gamma-1}\right]\left[1+(\gamma-2)\left(\frac{\erad}{\theta}\right)^{\gamma-1}\right],
\end{equation}
may be negative, depending on the image parity (which is not used). The magnification is plotted for different values of $\gamma$ in \cref{fig:image-structure}. The magnification diverges as $\theta\rightarrow\erad$, or similarly as $\beta\rightarrow 0$, where an infinitely magnified circular image is formed at $\erad$. For profiles with $\gamma<2$ the magnification also diverges at
\begin{equation}
      \label{eq:caustic}
	\theta = \erad(2-\gamma)^{1/(\gamma-1)}.
\end{equation}
The third, central image is located inside this angular radius.

 \begin{figure*}
 	\includegraphics[width=0.9\textwidth]{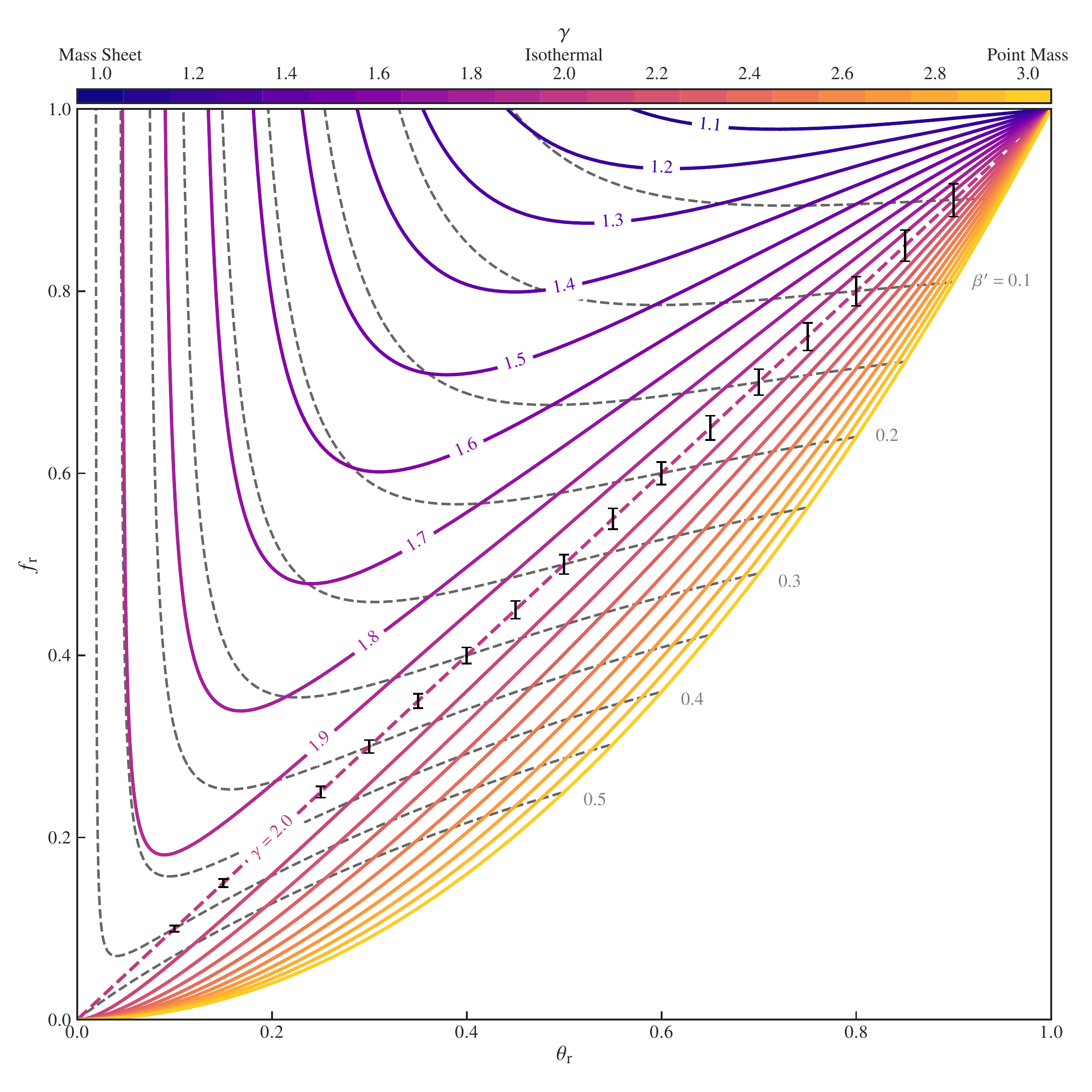}
 	\caption{\label{fig:position-ratio-flux-ratio-gamma} Contours of $\gamma$ and $\beta'$ as a function of $\flra$ and $\imra$. The dashed line shows the isothermal profile $\gamma=2$ where $\flra=\imra$. Error bars around the isothermal slope are calculated from \cref{eq:sigma-flux-ratio}, for $S=100$, and give an indication of the constraint on $\gamma$ from the observables. For example, at $\imra=0.8$ the error bars suggest we should achieve $\sg=0.1$, for $S=100$. It becomes easier to constrain $\gamma$ as $\imra$ decreases due to the shrinking $\flra$ error bars but also due to the increased spacing between contours of $\gamma$. Contours of $\beta'$ are found by solving \cref{eq:beta-dashed} with the value of $\gamma$ given by \cref{eq:flux-ratio}.}
 \end{figure*}
By combining \cref{eq:flra,eq:magnification} we have
\begin{equation}
\flra = \left|\left[\frac{1-\erad^{\gamma-1}\theta_2^{1-\gamma}}{1-\erad^{\gamma-1}\theta_1^{1-\gamma}}\right]
\left[\frac{1+(\gamma-2)\erad^{\gamma-1}\theta_2^{1-\gamma}}{1+(\gamma-2)\erad^{\gamma-1}\theta_1^{1-\gamma}}\right]\right|.
\end{equation}
Using \cref{eq:einstein-radius} we can eliminate $\erad$ and by substituting $\imra$ for the position ratio we finally obtain
\begin{equation}
\label{eq:flux-ratio}
\flra = \imra\left|\frac{\imra^{2-\gamma}+(\gamma-2)\imra+\gamma-1}{(\gamma-1)\imra^{2-\gamma} + (\gamma-2)\imra^{1-\gamma}+1}\right|.
\end{equation}

\cref {eq:flux-ratio} is the key equation in this paper. The equation may be numerically inverted to provide the relation between $\gamma$ and the two observables $\flra$ and $\imra$, demonstrating that a measurement of the flux ratio and position ratio for the two images provides a measurement of $\gamma$. This result is plotted in \cref{fig:position-ratio-flux-ratio-gamma}. We note that $\imra=\flra$ yields $\gamma=2$. This is evident by inserting $\gamma=2$ into \cref {eq:flux-ratio}, and is as expected for the isothermal sphere. Since $\beta'$ depends only on $\gamma$ and $\imra$ (see \cref{eq:beta-dashed}) it is also uniquely determined by $\flra$ and $\imra$. Contours of $\beta'$ are plotted in \cref{fig:position-ratio-flux-ratio-gamma} as dashed lines. The use of the coordinates $\flra$ and $\imra$ has eliminated the two system parameters $\erad$ and $f_\mathrm{S}$, which are irrelevant to the determination of $\gamma$.

It is significant that for $\gamma<2$ the contours of $\gamma$ become vertical at small values of $\imra$. This corresponds to the point where the inner image disappears, given by \cref{eq:caustic}, which is the same as setting the denominator of \cref {eq:flux-ratio} to zero. As an example, for $\gamma=1.5$ we find this condition holds when $\imra=3-2\sqrt{2}$.
 
\cref{fig:position-ratio-flux-ratio-gamma} makes clear the importance of flux information for measuring the mass profile. Measurement of the image positions alone provides only a degenerate combination of $\gamma$ and $\beta'$, equivalent to a vertical slice in \cref{fig:position-ratio-flux-ratio-gamma}. The addition of fluxes yields a single combination of $\gamma$ and $\beta'$, equivalent to a point in the figure. As is also evident from the figure, the uncertainties on $\gamma$ and $\beta'$ depend on the uncertainties on $\flra$ and $\imra$, as well as on the actual values of these two parameters, i.e. the location in the figure, since the spacing of the contours varies across the plane. 
Systems approaching a complete ring, where $\imra,\flra\rightarrow1$, can have any value of $\gamma$ and reproduce the same observables, leaving $\gamma$ unconstrained when $\beta'\rightarrow0$.

\subsection{Observational uncertainties}
\label{sec:noisemodel}

We use a simple model for the noise in the two images, that is tractable analytically, and that is a good approximation to the actual situation in images of interest, for example \textit{Hubble Space Telescope} (\textit{HST}) images of the BELLS GALLERY sample of \citet{Shu2016}. The main result in the current paper is the calculation of the uncertainty on the measured value of $\gamma$ as a function of three quantities: the combined S/N in the two images $S$, the position ratio $\imra$, and the value of $\gamma$ itself. Since the results are scaled to $S$ the only requirement on the noise model is to correctly represent the ratio of the S/N in the two images.

There will be four main contributions to the uncertainty in a particular image: i) detector noise (read noise and dark current); ii) sky background counts; iii) image counts from the source; and iv) counts at the location of the image from the lensing galaxy, that will have been subtracted. For the first two terms the contribution to the variance is proportional to the size of the image, which is proportional to the flux, because lensing preserves surface brightness. The third term is also proportional to the flux. For the fourth term, although the variance is again proportional to the size of the image, the constant of proportionality will be different for the two images, because the surface brightness of the lensing galaxy will be different at the two image locations. We assume simply that the variance in each image is proportional to the flux in the image. This will be a good approximation when the image counts from the source are greater than the counts from the lensing galaxy. This is true for example for the BELLS GALLERY sample. The model will also be generally valid when the contributions from the lensing galaxy at the locations of the two images are not too different. This will be true for larger values of the position ratio, $\imra\sim 1$. When the position ratio is small, and the lens galaxy is bright relative to the lensed source, the noise model will be less good. Configurations of this sort are less interesting in practise, because the magnifications are small, so cases with large S/N that are the most useful will be rare. The total magnification, summed for the two images, for the case $\gamma=2$ is given by
\begin{equation}
	\mu_\mathrm{tot}=\frac{2(1+\imra)}{1-\imra},
\end{equation}
and so for $\imra \leq 0.4$, $\mu_\mathrm{tot} \leq 14/3$.

\subsection{Uncertainty on the flux ratio}
\label{sec:uncertainty-on-the-flux-ratio}

The noise model is that the variance in an individual image is proportional to the flux, i.e. $\sigma_f^2 = af$ where $a$ is a constant. The summed signal to noise ratio $S$ in our two-image observation is then
\begin{equation}
\label{eq:snr}
S = \frac{f_1 + f_2}{\sqrt{af_1 + af_2}}.
\end{equation}
For a quantity $y$ which is a function of more than one parameter $y=f(x_1,...,x_n)$ the uncertainty on $y$, $\sigma_y$, is given by
\begin{equation}
\label{eq:general-error-propogation}
	\sigma_y^2=\sum_{i}^n \sigma_{x_i}^2\left|\pdv{f}{x_i}\right|^2,
\end{equation}
for small values of $\sigma_{x_i}$. This gives an expression for the uncertainty on the flux ratio
\begin{equation}
\label{eq:sigma-flra}
\sigma_\flra = \flra\left[\left(\frac{\sqrt{af_1}}{f_1}\right)^2 + \left(\frac{\sqrt{af_2}}{f_2}\right)^2\right]^\frac{1}{2}.
\end{equation}
Using \cref{eq:snr} to eliminate $a$ gives
\begin{equation}
\label{eq:sigma-flux-ratio}
\sigma_\flra = \frac{1}{S}\left(1+\flra\right)\sqrt{\flra}.
\end{equation}

\subsection{Uncertainty on the image position ratio}

Consider an image at position $\theta_k$. The radial size of the image is the source size $\reff$ multiplied by the radial magnification. For $\gamma=2$ the radial magnification is unity and at other values of $\gamma$ the correction to this is small enough to ignore here. If the image signal is normally distributed with variance $\reff^2$ around its true position and the position is estimated as the centroid of the image then
\begin{equation}
	\sigma_{\theta_k}^2=\frac{\reff^2}{S_k^2},
\end{equation}
where $S_k$ is the signal to noise ratio for that image, given by $S_k=f_k/\sqrt{af_k}$ (see \cref{sec:uncertainty-on-the-flux-ratio}). Again using \cref{eq:general-error-propogation} we add $\sigma_{\theta_k}$ for images 1 and 2 in quadrature
\begin{equation}
	\sigma_\imra = \imra \left[\left(\frac{\reff}{S_1\theta_1}\right)^2 + \left(\frac{\reff}{S_2\theta_2}\right)^2\right]^\frac{1}{2}.
\end{equation}
Considering the case where $\gamma=2$ we can use the fact that $\imra=\flra$ and $\theta_2=2\erad/(1+\imra)$ to simplify the above and obtain
\begin{equation}
	\label{eq:sigma-position-ratio}
	\sigma_\imra=\frac{1}{S}\frac{\reff}{\erad}\left[\frac{(\imra + 1)^3(\imra^3+1)}{4\imra}\right]^\frac{1}{2}.
\end{equation}
The term in brackets is $\sim 2$ for all but the smallest values of $\imra$.

\subsection{Uncertainty on the slope}
According to \cref{eq:flux-ratio} and \cref{fig:position-ratio-flux-ratio-gamma}, the slope can be determined via a measurement of both the position ratio and the flux ratio. Using \cref{eq:general-error-propogation}, the uncertainty on that measurement, $\sg$ can be estimated by
\begin{equation}
\label{eq:error-propogation}
\sg^2 = \left|\pdv{\gamma}{\flra}\right|^2\sigma_\flra^2 + \left|\pdv{\gamma}{\imra}\right|^2\sigma_\imra^2.
\end{equation}
\Cref{eq:flux-ratio} can be rewritten as a function $F$ of $\imra$, $\flra$ and $\gamma$, such that
\begin{equation}
F(\imra, \flra, \gamma)=0.
\end{equation}
We can then find the partial derivatives in \cref{eq:error-propogation} by the implicit function theorem,
\begin{equation}
\pdv{\gamma}{\imra}=-\frac{\pdv*{F}{\imra}}{\pdv*{F}{\gamma}},\quad \pdv{\gamma}{\flra}=-\frac{\pdv*{F}{\flra}}{\pdv*{F}{\gamma}}.
\end{equation}
The exact forms of the above are, in general, complicated; however, for an isothermal lens ($\gamma=2$) they become quite simple. We use the fact that $\imra=\flra$ and by evaluating the derivatives \cref{eq:error-propogation} becomes
\begin{equation}
\label{eq:sigma-isothermal-both-errors}
\sigma_{\mathrm{iso}} = \frac{2}{1-\flra^2} \sqrt{\sigma_\flra^2 + \sigma_\imra^2}.
\end{equation}
Now consider the relative contributions to the uncertainty from the measurements of flux ratio and position ratio. Using \cref{eq:sigma-flux-ratio,eq:sigma-position-ratio} the ratio of the uncertainties is
\begin{equation}
	\frac{\sigma_\imra}{\sigma_\flra}=\frac{\reff}{\erad}\left[\frac{(\imra^3 + 1)(\imra + 1)}{4\imra^2}\right]^\frac{1}{2},
\end{equation}
which is of the order $\reff/\erad$ at all except the smallest values of $\imra$. This shows that if the source is substantially smaller than the Einstein radius the uncertainty on the flux ratio dominates that of the position ratio and we can safely ignore the $\sigma_\imra^2$ term in \cref{eq:sigma-isothermal-both-errors}. Finally, then, for the uncertainty on the mass profile slope in a singular isothermal sphere, observed at a S/N of $S,$ we have
\begin{equation}
\label{eq:sigma-isothermal}
\sigma_{\mathrm{iso}} = \frac{2\sqrt{\flra}}{S\left(1-\flra\right)} = \frac{2\sqrt{\imra}}{S\left(1-\imra\right)}.
\end{equation}
Since galaxies are approximately isothermal, the accuracy with which $\gamma$  can be measured in a circular galaxy using strong lensing is encapsulated by this simple formula.

\begin{figure}
	\includegraphics[width=0.5\textwidth]{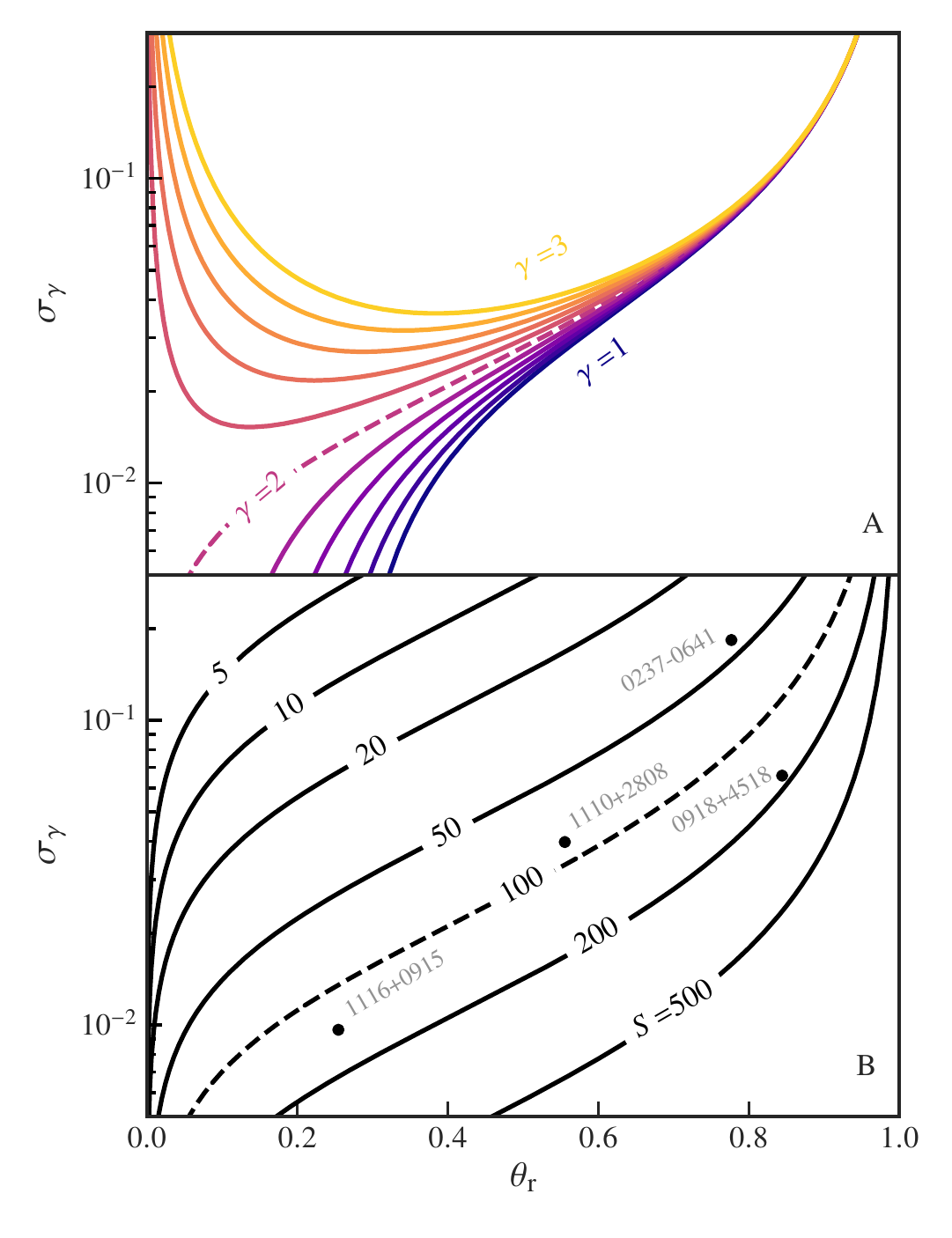}
	\caption{\label{fig:sigma-gamma} The uncertainty on the mass profile slope for: \textbf{A}. Different values of $\gamma$ with a fixed $S=100$. The dashed line shows the isothermal slope and lines are spaced by $\Delta\gamma=0.2$; \textbf{B.} Different values of $S$ with an isothermal slope. Values of $S$ are labelled. The curve for $S=100$ is compared with that derived from mock observations in \cref{sec:results}. The labelled points represent the expected $\sg$ for four systems in the BELLS GALLERY. For details of these systems see Table 2 in \citet{Shu2016}.}
\end{figure}

We have evaluated $\sg$ using \cref{eq:error-propogation}, and ignoring the $\sigma_\imra^2$ term, for a range of values of $\gamma$. From inspection of \cref{eq:sigma-flux-ratio} and \cref{eq:flux-ratio}, it can be seen that generally $\sg\propto 1/S$, with a complicated dependence on $\imra$ and $\gamma$, which simplifies to \cref{eq:sigma-isothermal} for $\gamma=2$. The results for different $\gamma$ and $S$ are plotted in \Cref{fig:sigma-gamma}. In the upper plot, we fix $S=100$, and plot curves of $\sg$ for different $\gamma$. In the lower plot, we fix $\gamma=2$ and vary $S$. In each panel, the dashed line corresponds to \cref{eq:sigma-isothermal}, for $S=100$. For an isothermal circular system with $S=100$ and $0.4<\imra<0.8$, \cref{eq:sigma-isothermal} yields $0.02<\sg<0.09$. The lower plot also includes some systems from the BELLS GALLERY (labelled in the figure) as illustrative examples. We selected four near-circular ($\varepsilon<0.15$), two-image systems and measured the S/N according to the method in \cref{sec:noise}.

The parameter $S$ separates out the dependence of $\sg$ on the quality of the observations, and reveals the dependence of $\sg$ on the lensing configuration. For a lens of particular $\gamma$, the only lens variable on which $\sg$ depends is $\imra$, and
$\sg$ decreases as $\imra$ decreases, or, equivalently, as the image separation increases. \Cref{fig:position-ratio-flux-ratio-gamma} illustrates the behaviour of the constraint in more detail. As $\imra$ decreases from unity, moving R to L in the figure, two effects combine to improve the constraint: $\gamma$ becomes less sensitive to changes in the observables, illustrated by the increased spacing in the contours; and the flux ratio $\flra$, which dominates the error budget, is more precisely measured, illustrated by the shrinking error bars. The result is an improving $\sg$ as the image separation increases for a constant signal to noise ratio, an effect which we will continue to see in the more complicated systems considered in later papers. To state this result another way, we can take the reciprocal of \cref{eq:sigma-isothermal} and use the image positions rather than the ratio to obtain
\begin{equation}
\frac{1}{\sigma_{\mathrm{iso}}}\propto\frac{\theta_2-\theta_1}{\sqrt{\theta_1\theta_2}}.
\end{equation}
This shows that the precision to which $\gamma$ is measured is
proportional to radial range $\theta_2-\theta_1$ divided by
the geometric mean of the image positions (recall that $\theta_1$ and $\theta_2$ are both positive quantities), which is very similar to the Einstein angle for small image separations (and is precisely the Einstein angle when $\gamma=3$).

In a population of real lenses $S$ is naturally a function of $\beta$. Moving the source away from the lens axis produces images of lower magnification and therefore smaller $S$ for the same level of noise. It is crucial to account for this in creating mock observations: fixing $S$ ensures that changes in $\sg$ across different image configurations are a function only of the physical ability of those configurations to constrain the mass distribution, rather than the quality to which they have been observed.

In summary, measuring the position ratio and flux ratio completely determines the slope in a circular power-law lens, irrespective of the Einstein angle or the structure in the source. Using the scalar magnification, the flux ratio can be written as a function of the image positions, the Einstein angle and the slope. The lens equation can then be used to eliminate the Einstein angle, resulting in an expression for the flux ratio which depends only on the position ratio and the mass profile slope, illustrated in \cref{fig:position-ratio-flux-ratio-gamma}. Estimating the uncertainty on the observables then gives an analytic expression for the uncertainty on the slope (\cref{eq:error-propogation}) which has a simple form for an isothermal lens (\cref{eq:sigma-isothermal}).

\section{SIMULATED OBSERVATIONS}
\label{sec:method}

In this section we detail the parameterisation of the lens and the source and describe the procedure for creating and inverting a simulated strong lensing observation. The purpose of the simulated observations is both to verify the previous analytic results, plotted in \cref{fig:sigma-gamma}, and to confirm the simulation and inversion methodology. In this way we will then be able to extend this work to more complicated problems using simulated observations alone. All the simulations use $\gamma=2$, and the noise is scaled such that the total S/N in the two images is $S=100$. If theory and simulation agree, we expect the measured uncertainties to lie along the dashed line plotted in the two panels in \cref{fig:sigma-gamma}.

We define a square image plane covering $6\,\mathrm{arcsec}\,\times\,6\,\mathrm{arcsec}$ with a pixel width of $0.04$~arcsec to mimic HST WFC3 observations. The image plane position vector at the centre of the $i$th pixel is $\vtheta_i$ and the surface brightness in this pixel is $s_i(\pvf)$ where $\pvf$ is the vector of lens and source parameters. We assume a transparent lens. We have not included the complication of convolution with a point spread function (PSF) in the modelling. In real images the effect of the PSF must be accounted for, but there is no difficulty in principle to correct for the PSF, and the lensing results will not be impacted provided the PSF HWHM is significantly smaller than the unlensed source effective radius, and the images are well sampled.

We define a function $I(\vbeta, \pvs)$ that gives the source plane surface brightness for a given source plane position vector $\vbeta$ and a set of source parameters $\pvs$. The lens equation then gives the surface brightness in the image plane
\begin{equation}
	\label{eq:surface-brightness}
	s_i = I\left[\vtheta_i - \valpha(\vtheta_i, \pvl), \pvs\right],
\end{equation}
where $\valpha$ is the vector deflection angle, itself a function of the lens parameters $\pvl$. The set of all image plane surface brightnesses comprises our model,
\begin{equation}
	\mathcal{M}\equiv\left\{s_i\right\}.
\end{equation}
The evaluation of $\mathcal{M}$ at a given $\pvf$ has two parts: first find the corresponding source plane coordinates for each image pixel, from the deflection angle; second compute the source surface brightness at those coordinates.

We now detail the parameterisation of the lens and the source. In fitting, as far as possible, it is important not to impose restrictions on the models, other than the assumption of a power law profile. Therefore although both the lens and the source as circular, in fitting we parameterise each as elliptical. For real data it is best to pixelise the source to avoid biasing the results  \citep{Nightingale2018}.

\subsection{Mass modelling}

We use a singular circular power-law galaxy for the lens, but fit a model of a singular power-law ellipsoid \citep[SPLE,][]{Tessore2015a} with the following parameters; a lensing strength $b$, a mass profile slope $\gamma$, an ellipticity $\varepsilon_\mathrm{L}$ and a position angle $\posa$. The ellipticity is defined as $\varepsilon=1-q$ where $q$ is the axis ratio (minor/major) of the mass distribution, and the position angle is the anti-clockwise angle from the $x$-axis to the semi-major axis of the mass distribution. The centroid of the mass distribution is assumed known. In a real observation the centroid of the mass would normally be taken as the centroid of the light of the lensing galaxy, which can usually be measured very precisely. Alternatively, the centroid could be parameterised.

The SPLE has homoeoidal, elliptical isodensity contours and its convergence is given by
\begin{equation}
	\kappa(\etheta) = \frac{3-\gamma}{2} \left(\frac{b}{\etheta}\right)^{\gamma-1},
\end{equation}
where $\etheta$ is an elliptical radius defined by
\begin{equation}
	\etheta^2 = q^2\theta_i^2 + \theta_j^2\:,
\end{equation}
where $i$ and $j$ are the major and minor axes respectively.
In other words, $\etheta$ is the semi-minor axis of the ellipse passing through $\vtheta=(\theta_i, \theta_j)$.

The lensing strength, $b$, is the elliptical analogue of the Einstein angle. Specifically it is the semi-minor axis of the ellipse where the average interior surface density is $\critd$. For a lens with lensing strength $b$ the equivalent circular lens has an Einstein angle $\erad=b/\sqrt{q}$. We use the solutions in \citet{Tessore2015a} to efficiently calculate $\valpha(\vtheta)$ for the SPLE.

In summary the mass model has four parameters, $b$, $\gamma$, $\varepsilon_\mathrm{L}$, and $\posa$.

\subsection{Source modelling}

We use a circular source with a S\'ersic profile to create the simulated observations, but fit an elliptical S\'ersic profile for the source surface brightness in the inversion. The elliptical S\'ersic profile is given by
\begin{equation}
I(\vbeta,\pvs) = I_0 \exp\left[-b_n\left(\frac{\beta_\varepsilon(\vbeta)}{\reff}\right)^{1/n_s}\right],
\end{equation}
where $I_0$ is the surface brightness at the centre of the source, given by $\vbeta_\mathrm{S}=\left(\beta_{\mathrm{S}_x},\beta_{\mathrm{S}_y}\right)$ and $\reff$ is the effective radius. $\beta_\varepsilon(\vbeta)$ is an elliptical radius in the source plane given by
\begin{equation}
	\beta_\varepsilon^2=q_\mathrm{S}^2\left(\beta_x-\beta_{\mathrm{S}_x}\right)^2+\left(\beta_y-\beta_{\mathrm{S}_y}\right)^2,
\end{equation}
where $q_\mathrm{S}$ is the source axis ratio, defined in the same way as the lens axis ratio above, and $\beta_\varepsilon$ is appropriately transformed depending on the source position angle $\poss$. The source ellipticity is given by $\varepsilon_\mathrm{S}=1-q_\mathrm{S}$. An expression for the constant $b_n$, chosen such that $\reff$ is the half-light radius of the galaxy, is provided by \citet{Ciotti1999}.

To correctly sample the source function near its centre we adaptively sub-pixelise the image plane as follows. \Cref{eq:surface-brightness} is evaluated a single time for each pixel. Each pixel is then assigned a level of sub-pixelisation proportional to its surface brightness in the first pass. The mean of the sub-pixel brightnesses then gives the final brightness in each pixel. Up to $100$ sub-pixels are used in the brightest pixels.

In summary the source model has seven parameters, $\beta_{\mathrm{S}_x}$, $\beta_{\mathrm{S}_y}$, $I_0$, $n_s$, $\reff$, $\varepsilon_\mathrm{S}$ and $\poss$.

\subsection{Addition of noise}
\label{sec:noise}

The process outlined so far produces a model image plane $\mathcal{M}$ for a given set of input parameters $\pvf$. We now use this model evaluation to produce a mock observation by adding noise. The noise model, described in \cref{sec:noisemodel}, requires that the variance in each of the images is proportional to the flux in the image. This can be achieved most simply by applying uniform noise across the image plane, since image size is proportional to flux. We would like to produce a set of simulated observations of the same $S$. Therefore we need a process for fixing $S$ in a consistent way across different simulated observations.

\begin{figure*}
	\includegraphics[width=0.9\textwidth]{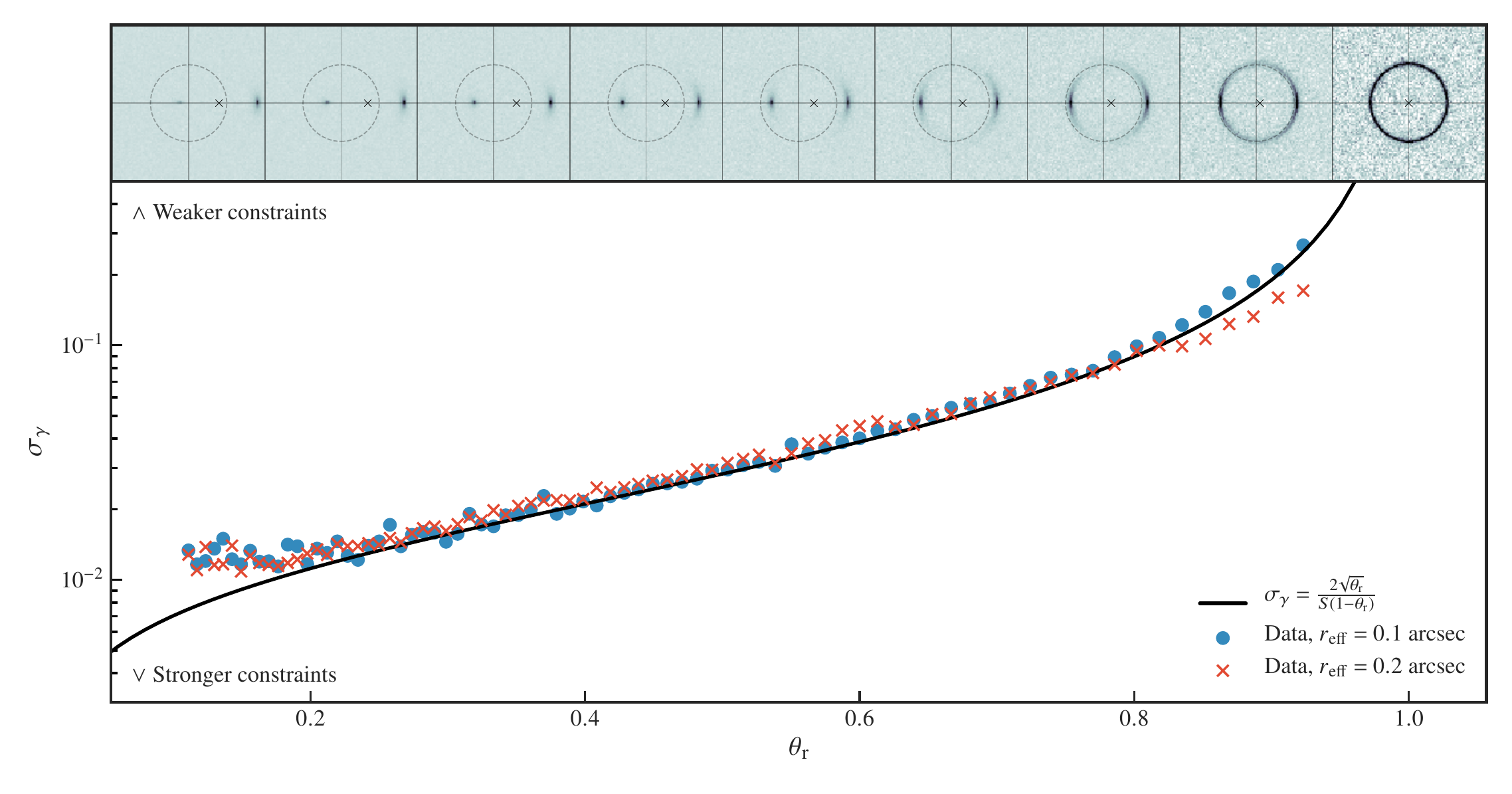}
	\caption{\label{fig:sigma-gamma-observation} The constraint on the mass profile slope $\sg$ from both sets of mock observations as a function of image position ratio, $\imra$. The solid curve is the predicted constraint from the analysis in \cref{sec:theory}, specifically \cref{eq:sigma-isothermal}, with $S=100$. The upper frames show the $\reff=0.1$ arcsec observations at the corresponding $\imra$. Source position and Einstein radii are plotted as crosses and dotted curves respectively.}
\end{figure*}

To do this we add noise $n_i$ to each pixel from a normal distribution with zero mean and variance $\sigmad^2$.
The new set of surface brightness values with added noise comprise our data $\mathcal{D}$:
\begin{equation}
	\mathcal{D}\equiv\{d_i\}=\{s_i+n_i\}.
\end{equation}
If the $n_i$ are distributed around zero with variance $\sigmad^2$ then the total signal to noise ratio in the simulated observation is
\begin{equation}
	S = \frac{\sum_i^N d_i}{\sigmad \sqrt{N}},
\end{equation}
where the sum is restricted to only include pixels that belong to the two lensed images. To find these pixels we convolve the image plane with a Gaussian kernel. The kernel standard deviation scales with source size, and we use $\sigma=0.08$ arcsec (2 pix.) for a source with $\reff=0.1$ arcsec. After convolving, image pixels are chosen as those with brightness above $2\sigmad$. We assign a masking variable $m_i$ where $m_i=1$ for an image pixel, $0$ otherwise.
The level of noise $\sigmad$ required to reach a target signal to noise ratio $S_\mathrm{T}$ is then given by
\begin{equation}
	\sigmad=\frac{\sum_i^N m_i d_i}{S_\mathrm{T} \sqrt{\sum_i^N m_i}}.
\end{equation}
The values of $m_i$ depend on $\sigmad$ so the $\sigmad$ which sets the correct $S_\mathrm{T}$ must be found iteratively.

To decide on a suitable value for the total S/N in our mock observations we measured $S$ in the $2\sigmad$ mask for $16$ images in the BELLS GALLERY sample \citep{Shu2016}, which were found to have $50<S<500$ with a mean $S=241$. Based on this we chose $S=100$ as representative of the typical quality of images available for fitting lens models.

\subsection{Parameter estimation}

Recall that our data $\mathcal{D}$ is a set of pixel values $\{d_i\}$ where each $d_i=s_i+n_i$ and each $n_i$ is drawn from $\mathcal{N}(0,\sigmad)$. With this assumption, the probability that we observe $d_i$ in a given pixel is
\begin{equation}
	\prob{d_i}{s_i} = \frac{1}{\sqrt{2\pi \sigmad^2}} \exp\left[-\frac{(d_i - s_i)^2}{2\sigmad^2}\right],
\end{equation}
where $s_i$ can be found by \cref{eq:surface-brightness} for a given $\pvf$. The product of these probabilities from all $N$ pixels
\begin{equation}
	\prob{\mathcal{D}}{\mathcal{M}} = \prod_{i}^{N} \prob{d_i}{s_i},
\end{equation}
gives the probability of observing $\mathcal{D}$ if the true model is $\mathcal{M}$. The model is only a function of the parameters $\pvf$ so we can write the above as
\begin{equation}
	\label{eq:data-likelihood}
	\prob{\mathcal{D}}{\pvf} = C \exp\left(-\frac{1}{2}\chi^2\right),
\end{equation}
where $C$ is some constant dependent only on $\sigmad$ and $\chi^2$ has the standard definition
\begin{equation}
	\chi^2 = \sum_{i}^{N} \left(\frac{d_i - s_i}{\sigmad}\right)^2.
\end{equation}
Using Bayes's theorem,
\begin{equation}
	\prob{\pvf}{\mathcal{D},\mathcal{I}} = \frac{\prob{\pvf}{\mathcal{I}}\prob{\mathcal{D}}{\pvf}}{\prior{\mathcal{D}}},
\end{equation}
we obtain the posterior probability of the model parameters as
\begin{equation}
	\label{eq:post}
	\prob{\pvf}{\mathcal{D},\mathcal{I}} \propto \prob{\pvf}{\mathcal{I}} \exp\left(-\frac{1}{2}\chi^2\right),
\end{equation}
where $\prob{\pvf}{\mathcal{I}}$ is the probability of the model having parameters $\pvf$ before any observation, also called the `prior' probability of $\pvf$. This is conditioned on prior information $\mathcal{I}$ which in this case is the physical range of the parameters. We adopt a prior that is uniform and proper in each parameter within the finite bounds specified by \cref{table:modelparams}.

We then marginalise over the parameters $\beta_{\mathrm{S}_x}$, $\beta_{\mathrm{S}_y}$, $I_0$, $n_s$, $\reff$, $\varepsilon_\mathrm{S}$, $\poss$, $b$, $\varepsilon_\mathrm{L}$, and $\posa$ to find $\prob{\gamma}{\mathcal{D},\mathcal{I}}$. We define the posterior uncertainty in $\gamma$, $\sg$, as the mean distance from the median of $\prob{\gamma}{\mathcal{D},\mathcal{I}}$ to its $16$th and $84$th percentiles. If $\prob{\gamma}{\mathcal{D},\mathcal{I}}$ is normally distributed then $\sg$ is equivalent to the $1\sigma$ uncertainty. Using $\sg$ to describe the constraint on $\gamma$ in this way is a convenient but incomplete summary of the posterior which will also feature correlations between the parameters. In later parts where we refer to `the constraint on the profile slope' or similar, $\sg$ as defined here is the specific quantity we are referring to.

\subsection{Implementation and posterior sampling}

For a given mock observation we constrain the parameters using a Markov chain Monte Carlo (MCMC) sampler. We use the affine-invariant ensemble sampler introduced by \citet{Goodman2010} and implemented in the \texttt{emcee} Python package by \citet{Foreman-Mackey2013}. The affine-invariance of the sampler makes it very useful for the highly correlated posterior distributions present in lensing problems. It also scales well in parallel, further improving convergence time. 
The sampler uses the log-posterior
\begin{equation}
	\log\prob{\pvf}{\mathcal{D},\mathcal{I}} = \log\prob{\pvf}{\mathcal{I}} -\frac{1}{2} \chi^2.
\end{equation}
When sampling, we swap the parameters $\varepsilon$ and $\posa$ for $\varepsilon_x$ and $\varepsilon_y$, defined such that
\begin{align}
	& \varepsilon = \sqrt{\varepsilon_x + \varepsilon_y},\\
	& \posa = \tan[-1](\varepsilon_y / \varepsilon_x),
\end{align}
which accounts for the unconstrained position angle when the lens is circular \citep{Hezaveh2016}. We take the same approach with the source ellipticity and position angle.

To ensure as short a burn-in time as possible we initialise the walkers in a Gaussian ball around the true parameter values which we already know for our simulated observations. We could opt to initialise the walkers in a uniform distribution across the space defined by \cref{table:modelparams} and achieve the same results, albeit with a longer convergence time.  After an initial burn-in phase with a small number of walkers ($\sim 50$) the number of walkers is increased ($\sim300$), using the final positions of the burn-in walkers as seed locations. Again, this is done in the interest of speed. At the end of the sampling run, walkers that did not converge to the main ensemble are removed, found automatically using a clustering algorithm. Typically $\lesssim1$ per cent of the walkers are removed in this way.

\section{Results}
\label{sec:results}

\subsection{Summary of simulated observations}
\label{sec:modelssummary}

The simulated observations all have circular, isothermal lenses with $\erad=1.0$ arcsec. The sources are also circular, with $n_s=2$, and central surface brightness (which is arbitrary) set to unity.  Two sets of simulated observations were created, one with the source effective radius $\reff=0.1$ arcsec, and one with $\reff=0.2$ arcsec. For each set the source position was moved along the $x$ axis over the range $0\leq\beta_{\mathrm{S}_x}\leq0.8$ arcsec in steps of 0.01 arcsec. The relation between $\imra$, $\flra$ and $\beta$ for an isothermal lens is
\begin{equation}
	\imra = \flra = \frac{\erad - \beta}{\erad + \beta},
\end{equation}
so the simulated observations cover the range $0.11\leq \imra \leq 1.0$.
All the parameters, and their priors, are summarised in \cref{table:modelparams}.
Pictures of the models, for a range of $\imra$ are provided in the top panel of \cref{fig:sigma-gamma-observation}, for the case of $\reff=0.1$ arcsec.

\begin{figure}
	\includegraphics[width=0.45\textwidth]{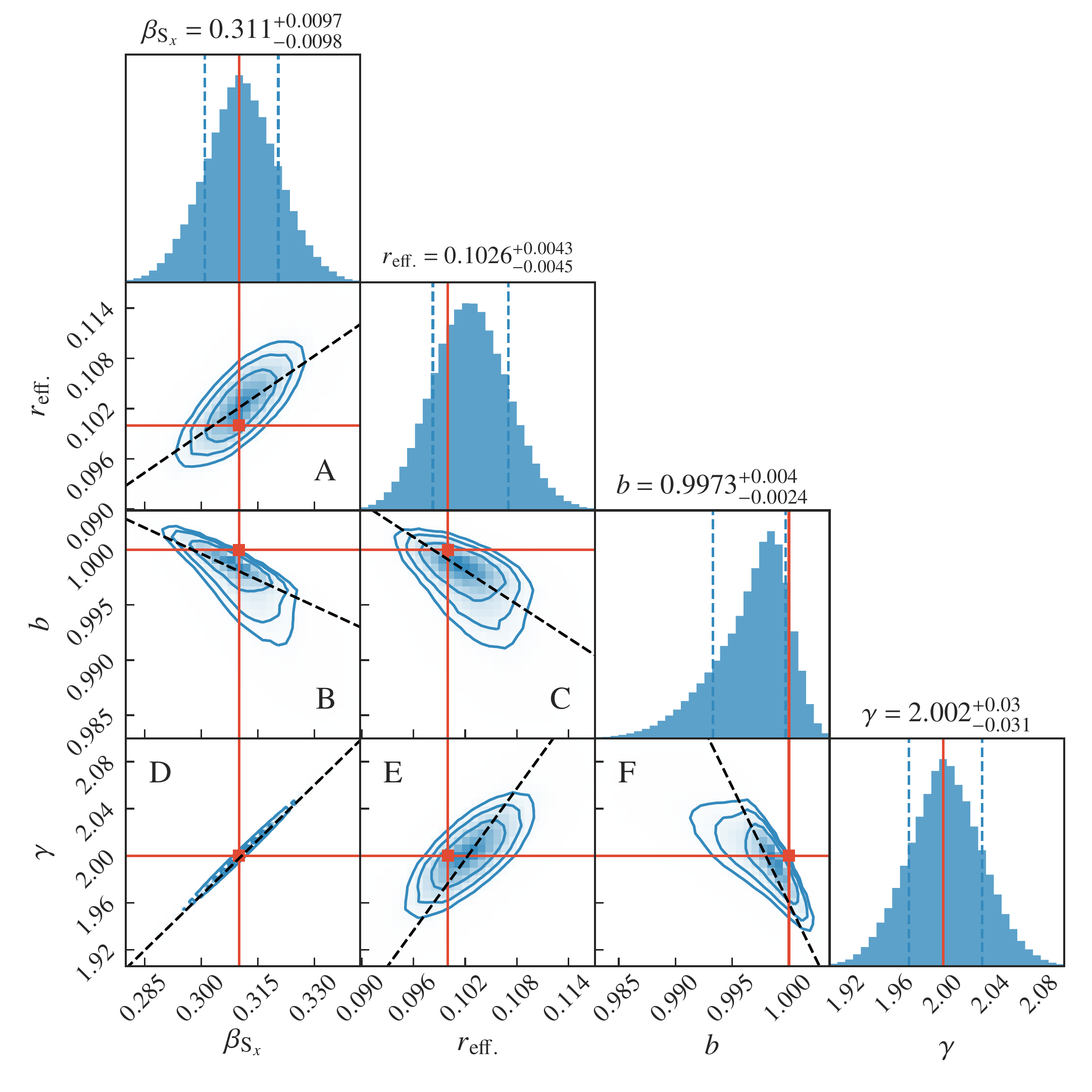}
	\caption{\label{fig:corner-plot}The posterior probability density function in four of the eleven parameters for a system with $\imra=0.51$ or $\beta=0.31$ arcsec. Contours indicate the $68\%$, $95\%$ and $99.7\%$ posterior density credible region. On the marginal distributions the dashed lines indicate the $68\%$ posterior credible region, quantified above each. The true values used to create the observation are marked in red on each plot. The black dashed lines are the predicted correlations between the lens and source parameters. The gradients for these are given in \cref{sec:correlations} and an intercept is added such that each line passes through the posterior mode. For the sake of clarity we omit the posteriors for $\beta_{\mathrm{S}_y}$, $I_0$, $n_s$, $\varepsilon_{\mathrm{S}_x}$, $\varepsilon_{\mathrm{S}_y}$, $\varepsilon_x$ and $\varepsilon_y$. They are all well constrained and uncorrelated with the parameters above.}
\end{figure}

\subsection{Constraints on the mass profile slope}

\begin{table}
	\begin{tabular}{llll}
	&	\textbf{Parameter} & \textbf{Model} & \textbf{Prior limits}\\
		\hline
	&		$b$ arcsec & $1$ & $0<b\leq10$ \\
Lens &		$\gamma$ & 2 & $1\leq\gamma<3$ \\
	&		$\varepsilon_x$ & 0 & $-1\leq\varepsilon_x<1$ \\
	&		$\varepsilon_y$ & 0 & $0\leq\varepsilon_y<1$ \\
		\hline
	&		$\{\beta_{\mathrm{S}_x}, \beta_{\mathrm{S}_y}\}$ arcsec & $\{0.0-0.8,0.0\}$ & $|\vbeta_\mathrm{S}|<1$ \\
Source &	$I_0$ & $1$  & $0<I_0\leq10$\\
	&		$n_s$ & $2$ & $0<n_s\leq6$\\
	&		$\reff$ & 0.1, 0.2 & $0<\reff<1$ \\
	&		$\varepsilon_\mathrm{S_x}$ & 0 & $-1\leq\varepsilon_\mathrm{S_x}<1$ \\
	&		$\varepsilon_\mathrm{S_y}$ & 0 & $0\leq\varepsilon_\mathrm{S_y}<1$ \\
		\hline
	&		S & $100$ & \\
		\hline
	\end{tabular}
	\caption{\label{table:modelparams}The parameters, and their priors, for the synthetic observations. The priors on the ellipticity parameters are equivalent to restricting the lens or source position angle to the range $0\leq \phi <\pi$.}
\end{table}

In \cref{fig:sigma-gamma-observation} we plot the measured value of $\sg$ for all the simulated observations, together with the theoretical curve according to \cref{eq:sigma-isothermal}. The constraints on the mass profile slope measured from the simulations match the predicted curve very closely, for both values of $\reff$. This result confirms that the theoretical analysis captures all that is relevant in measuring $\gamma$. The earlier analysis made no assumptions about the source structure, yet we have correctly predicted $\sg$ for extended sources. This shows that the theoretical analysis should apply equally to any source brightness distribution, and the unknown structure of the source \---\ if properly treated (see below) \---\ has no impact on the precision of the measurement of $\gamma$. The result also verifies the simulation methodology, including the method for measuring $S$ in the images.

The size of the source has no significant effect on the constraint as long as $\reff\ll\erad$, as is true for both cases above. By assuming that $\sigma_\flra$ dominates over $\sigma_\imra$ in deriving \cref{eq:sigma-isothermal} we removed any dependence of $\sg$ on $\reff$ and the data shows this assumption to be valid. In our mock observations the true (isothermal) slope is found $64.9\%$ of the time to within $1\sigma$ and $94.1\%$ of the time to within $2\sigma$, confirming that the results are unbiased. For real data a S\'ersic profile will only be an approximation to the true source profile, and as noted earlier to avoid biased results a pixelised source should be used \citep{Nightingale2018}.

At $\imra\rightarrow1$ we should in theory see $\sg\rightarrow\infty$ as there is only one infinitely magnified image at $\theta=\erad$ (see upper right of \cref{fig:position-ratio-flux-ratio-gamma}). However there is some evidence that the constraint for $\reff=0.2\,\mathrm{arcsec}$ is slightly better than the theoretical prediction for large values of $\imra$. We attribute this to the fact that in the simulated observations the Einstein ring has a finite width and the small difference in $\imra$ and $\flra$ between pixels across the extension of the ring is enough to provide a weak constraint on $\gamma$. At the other extreme, as $\imra\rightarrow0$, both sets of data start to deviate from the prediction. At $\imra<0.2$ the term in brackets in \cref{eq:sigma-position-ratio} begins to increase and our assumption that the position ratio uncertainty does not contribute to $\sg$ starts to break down.

\subsection{Parameter constraints and correlations}

Analysis of the correlations between the parameters of the fit provides further insight into the lensing properties of the singular power-law galaxy. The form of the posterior probability density functions (PDFs) is similar across the range of systems, containing a complicated set of correlations. \Cref{fig:corner-plot} shows the PDF for four parameters that display interrelated correlations. Additionally the parameters $n_s$ and $I_0$ are correlated. This is a generic feature in fitting S\'ersic profiles to galaxies, and is unrelated to determination of the lensing parameters.

The correlations plotted may all be understood through the earlier theoretical analysis. We have computed the theoretical slopes of the correlations for the six panels labelled A$-$F in the figure, plotted as dashed lines, and their derivation is explained below. All the computed correlations match the results of the simulations well. The correlation between $\beta_{\mathrm{S}_x}$ and $\gamma$ is the degeneracy between these parameters for fixed $\imra$, corresponding to a vertical line in \cref{fig:position-ratio-flux-ratio-gamma}. The degeneracy is broken by the flux information, and the uncertainty along the line is set by the uncertainty on the flux ratio. Although the correlation is extremely narrow, the axis ratio of the correlation relates to the ratio of the uncertainty on $\imra$ to the uncertainty on $\flra$. For a larger source the correlation is broader.

To derive the slope of the correlation it is sufficient to assume that the positions are measured perfectly. Then we simply differentiate \cref{eq:source-position} with respect to $\gamma$, assuming $\theta_1$ and $\theta_2$ are constant. Since the model is isothermal, we insert $\gamma=2$ into the resulting expression, and substitute $b=\erad=(\theta_1+\theta_2)/2$ from \cref{eq:einstein-radius} to yield the simple relation
\begin{equation}
\dv{\gamma}{\beta}=-\frac{2}{b \log \imra},
\end{equation}
which is plotted in panel D in \cref{fig:corner-plot}. In a similar way we differentiate \cref{eq:einstein-radius} with respect to $\gamma$ to derive the correlation between $\gamma$ and $b$

\begin{equation}
\dv{\gamma}{b}= \frac{2}{b\log\left(b^2-\beta^2\right)}\,,
\end{equation}
which is plotted in panel F. Application of the chain rule leads to the correlation between $b$ and $\beta$, plotted in panel B.

In frames A, C and E in \cref{fig:corner-plot}  the inferred radius of the source is correlated with the system parameters $\beta$, $b$, and $\gamma$. This arises because of the dependence of magnification on $\gamma$ for fixed position $\theta$, \cref{eq:magnification} . Since the inferred source radius is inversely proportional to the square root of the absolute value of the magnification $\reff \propto \mu^{-1/2}$, then 
\begin{equation}
\label{eq:magnification-correlation}
\dv{\reff}{\mu} = -\frac{\reff}{2\mu}.
\end{equation}
From \cref{eq:magnification} we can compute $\dd{\mu}/\dd{\gamma}$. Combining these two relations with the derivatives of \cref{eq:einstein-radius,eq:source-position} we can compute all the correlations in frames A, C and E. Where $\mu$ appears in an expression for a correlation, as in these three panels, we compute the gradient at both images and take the mean weighted by image size, proportional to image position for an isothermal lens. For the explicit route to each of the gradients used in the plot see \cref{sec:correlations}.

\section{Conclusions}

We have presented an analysis of how observations of an extended background source lensed by a singular circular power-law lens can be used to measure the slope $\gamma$ of the density distribution $\rho\propto r^{-\gamma}$ in the lens. Using just the positions of the two images provides only a degenerate combination of $\gamma$ and the source position $\beta$. The flux information breaks this degeneracy. We showed that if the lens is isothermal, with $\gamma=2$, the precision with which $\gamma$ may be measured is given by a simple analytic expression. This precision improves with decreasing image position ratio, or increasing image separation, and increasing signal to noise ratio.

The theoretical analysis was based simply on the measurement of two quantities, the position ratio and the flux ratio, and made no assumption about the source surface-brightness profile. Nevertheless the results were confirmed to apply in the case of a full inversion of lensed images for an extended source. We created synthetic observations for a circular isothermal power-law lens, and a circular source, but we fit with elliptical models for both the lens and source. We recovered unbiased values of the 11 input parameters of the lens and the source, and the results for $\sigma_\gamma$ as a function of $\imra$ match the theoretical curve accurately. The fact that the results of the full inversion, fitting the surface-brightness distribution, agree with the theoretical analysis, which made no assumptions about the source surface-brightness profile, shows that fitting the surface brightness profile in a real case, using e.g. the method of \citet{Nightingale2018}, utilises the full positional and flux information in determining the mass slope. In effect a complicated galaxy light profile may be considered as a set of compact sources, each contributing to the measurement of $\gamma$.

These results, especially the strong agreement between the theoretical analysis and the synthetic observations, validate our inversion method and our method for setting the total signal to noise ratio. This is crucial to the work in the rest of this series where purely analytic results will not be possible and we will have to rely on the synthetic observations alone. Having established a theory for the constraints on the profile slope in the simplest case, the rest of the series will address primarily two issues: the radial range of the galaxy over which the constraint on the slope applies, and the strength of constraint available in elliptical lenses.

\section*{Acknowledgements}
We are grateful to Simon Dye for discussions on aspects of this work, to Yiping Shu for sending us images from the BELLS GALLERY sample, and to the Imperial College Research Computing Service for HPC resources and support. SJW thanks Tom Broadhurst for a conversation long ago, before CMO'R went to school, that sparked interest in this topic. CMO'R is supported by an STFC Studentship.




\bibliographystyle{mnras}
\bibliography{bibliography} 


\appendix

\section{Correlations in Power Law Profiles}
\label{sec:correlations}
We have six correlations between four parameters; $\beta$, $\reff$, $b$ and $\gamma$. In \cref{sec:theory} we derived three equations which relate these parameters. Combining their derivatives can give us each of the six correlations in \cref{fig:corner-plot}. We define the derivatives we need as follows.
\begin{enumerate}
	\item From \cref{eq:einstein-radius} with $b=\erad$
	\begin{equation}
		D_1 = \dv{b}{\gamma} = \dv{}{\gamma}\left[\left(\frac{\theta_1 + \theta_2}{\theta_1^{2-\gamma} + \theta_2^{2-\gamma}}\right)^\frac{1}{\gamma-1}\right].
	\end{equation}
	\item From \cref{eq:source-position}
	\begin{equation}
		D_2 = \dv{\beta}{\gamma} = \dv{}{\gamma}\left(\frac{\theta_2^{\gamma-1} - \theta_1^{\gamma-1}}{\theta_1^{\gamma-2} + \theta_2^{\gamma-2}}\right).
	\end{equation}
	\item From \cref{eq:magnification}
	\begin{equation}
		D_3 = \dv{\mu}{\gamma} = \dv{}{\gamma}\left\{\left[1-\left(\frac{\erad}{\theta}\right)^{\gamma-1}\right]^{-1}\left[1+(\gamma-2)\left(\frac{\erad}{\theta}\right)^{\gamma-1}\right]^{-1}\right\}.
	\end{equation}
	\item By assuming that $\reff\propto 1/\sqrt{\mu}$
	\begin{equation}
		D_4 = \dv{\reff}{\mu} = \frac{\reff}{2\mu}
	\end{equation}
\end{enumerate}
In each frame we then assume the correlation lies on a straight line with the gradient given by the derivative that combines those two parameters. The relevant gradients for each  of the highlighted frames in \cref{fig:corner-plot} are then:
\begin{alignat}{4}
	&\mathrm{Frame\, A.}\quad&&\dv{\reff}{\beta}  \quad&&=\dv{\reff}{\mu}\dv{\mu}{\gamma}\dv{\gamma}{\beta} 	\quad&&= \frac{D_3 D_4}{D_2},\\[7pt]
	&\mathrm{Frame\, B.}\quad&&\dv{b}{\beta}      \quad&&=\dv{b}{\gamma}\dv{\gamma}{\beta} 					&&= \frac{D_1}{D_2},\\[7pt]
	&\mathrm{Frame\, C.}\quad&&\dv{b}{\reff}      \quad&&=\dv{b}{\gamma}\dv{\gamma}{\mu}\dv{\mu}{\reff}		&&= \frac{D_1}{D_3 D_4},\\[7pt]
	&\mathrm{Frame\, D.}\quad&&\dv{\gamma}{\beta} \quad&&=\frac{1}{D_2},										&&\\[7pt]
	&\mathrm{Frame\, E.}\quad&&\dv{\gamma}{\reff} \quad&&=\dv{\gamma}{\mu}\dv{\mu}{\reff} 					&&= \frac{1}{D_3 D_4},\\[7pt]
	&\mathrm{Frame\, F.}\quad&&\dv{\gamma}{b}     \quad&&=\frac{1}{D_1}.										&&
\end{alignat}
Exact expressions for the derivatives can be found by symbolic computation software and numerically evaluated for the specific image plane configuration at hand.


\bsp	
\label{lastpage}
\end{document}